# High-efficiency and broadband electro-optic frequency combs enabled by coupled micro-resonators

Yaowen Hu[1,2,*], Mengjie Yu[1,*], Brandon Buscaino[3], Neil Sinclair[1], Di Zhu[1], Rebecca Cheng[1], Amirhassan Shams-Ansari[1], Linbo Shao[1], Mian Zhang[4], Joseph M. Kahn[3], and Marko Lončar[1,†]

[1]*John A. Paulson School of Engineering and Applied Sciences, Harvard University, Cambridge, MA 02138, USA*
[2]*Department of Physics, Harvard University, Cambridge, MA 02138, USA*
[3]*Edward L. Ginzton Laboratory, Department of Electrical Engineering, Stanford University, Stanford, CA 94305, USA*
[4]*HyperLight Corporation, 501 Massachusetts Ave, Cambridge, MA 02139, USA*
[*]*These authors contributed equally*
[†]*Correspondence to: loncar@seas.harvard.edu*

**Developments in integrated photonics have led to stable, compact, and broadband comb generators that support a wide range of applications. Current on-chip comb generators, however, are still limited by low optical pump-to-comb conversion efficiencies. Here, we demonstrate an integrated electro-optic frequency comb with a conversion efficiency of 30% and an optical bandwidth of 132 nm, featuring a 100-times higher conversion efficiency and 2.2-times broader optical bandwidth compared with previous state-of-the-art integrated electro-optic combs. We further show that, enabled by the high efficiency, the device acts as an on-chip femtosecond pulse source (336 fs pulse duration), which is important for applications in nonlinear optics, sensing, and computing. As an example, in the ultra-fast and high-power regime, we demonstrate the observation of a combined EO-$\chi^{(3)}$ nonlinear frequency comb. Our device paves the way for practical optical frequency comb generators enabling energy-efficient computing, communication, and metrology, and provides a platform to investigate new regimes of optical physics that simultaneously involve multiple nonlinearities.**

Integrating optical frequency comb (OFC) generators into chip-based devices opens opportunities for robust, compact, portable, and scalable comb sources enabling a wide range of applications, including communication [1], ranging [2], spectroscopy [3], frequency metrology [4], optical computing [5,6], and quantum information [7,8]. To date, the majority of on-chip OFCs originate from continuous-wave (pump) light driving resonantly enhanced $\chi^{(3)}$ optical nonlinearities in the anomalous group velocity dispersion (GVD) regime, producing Kerr frequency combs [9]. While these OFCs can feature attractive properties, such as octave-spanning bandwidths at THz-range comb line spacings, Kerr combs are limited by low pump-to-comb conversion efficiencies of a few percent, owing to the special mode-locking regime in which the pump is detuned far from the cavity resonance [10,11]. Another approach for improving the efficiency of Kerr combs has recently emerged that relies on dark-pulses [12–14]. Dark-pulse Kerr combs are generated in the normal GVD regime and can increase the conversion efficiency to ~10-50%. However, this comes at the expense of broader temporal widths, which renders them unsuitable for ultra-fast (e.g., femtosecond) pulse generation. In addition, dark-pulse combs are limited to fewer (~100) comb lines and large line spacings of >100 GHz. Finally, characteristic of all Kerr frequency combs is the existence of a pump threshold that is determined by both the Kerr nonlinearity and quality factor ($Q$) of the resonator. This results in a nonlinear dependence between the comb and pump powers, which leads to saturation effects and limits the absolute comb power.

Electro-optic (EO) modulation provides an attractive alternative for OFC generation [15–21]. Their electrical controllability provides not only versatility, but also excellent comb stability and phase coherence compared to on-chip OFCs generated using $\chi^{(3)}$ optical nonlinearities. EO combs based on conventional (travelling-wave) modulators feature a high pump-to-comb conversion efficiency, but their bandwidths span only a few nanometers due to the weak frequency mode interaction during a single pass through the modulator [15,16]. Wider optical combs can be generated using cavity-based EO combs in which light passes through a phase modulator multiple times while circulating inside an optical microresonator (Fig. 1a) [17–22] Recent progress on thin-film lithium niobate (TFLN) has enabled on-chip EO combs with a record-high optical bandwidth of ~80 nm [21], an order of magnitude higher than previous EO combs. However, the comb-conversion efficiency of this single-resonator EO comb generation is limited to only ~0.3%. Such a low conversion efficiency originates from a strongly under-coupled "hot" cavity resonance (<1% extinction ratio) when the generator is driven by a strong microwave tone. As a result, most of the light is not coupled into the resonator and is transmitted through the bus waveguide without entering the cavity (Fig. 1c).

Here, we address the low conversion efficiency of cavity-based EO comb and experimentally demonstrate an on-chip EO comb with a line spacing of 30.925 GHz, a pump-to-comb conversion efficiency of 30%, and a broad comb bandwidth of 132 nm. This is enabled by two mutually coupled resonators (Fig. 1d), realized in the TFLN platform [23]. A small resonator (cavity 1) is used to over-couple only the pump mode of a racetrack cavity (cavity 2 for comb generation) while rejecting the other non-pump modes (Fig. 1e). This results in a critically coupled two-resonator device when the microwave modulation is on (Fig. 1f) and increases the conversion efficiency as theoretically predicted [11,24,25]. We use the tight-binding model, previously developed for frequency crystals [26], as well as the generalized critical coupling (GCC) condition, developed for electro-optic frequency shifters and beam splitters [27], to model our two-resonator EO comb generator. Importantly, our theoretical approach allows the derivation of an analytical solution for the system and can be extended to multiple coupled resonator devices that may provide further novel functionalities.

The origin of the low conversion efficiency for single-resonator EO combs is the strong effective loss rate $\kappa_{MW}$ for the pump mode, induced by the microwave modulation, which extracts power from the pump mode into other comb lines. The loss rate $\kappa_{MW}$ is (see Supplementary Materials):

$$\kappa_{MW} = \kappa \left( \sqrt{1 + \frac{4\Omega^2}{\kappa^2}} - 1 \right) \tag{1}$$

where $\Omega$ is the mode-coupling rate between resonator modes separated by the FSR, and is proportional to the microwave voltage, while $\kappa = \kappa_e + \kappa_i$ is the cavity loss rate, with $\kappa_i$ and $\kappa_e$ being the intrinsic loss and coupling rate to bus waveguide, respectively. Here $\kappa_{MW}$ can be orders of magnitude higher than $\kappa_e$ and $\kappa_i$ (e.g., $\kappa_{MW} \sim 10$ GHz, $\kappa_e \sim \kappa_i \sim 100$ MHz for TFLN). Broadband EO combs require a strong microwave driving power (large $\Omega$) which leads to a large $\kappa_{MW}$. This, however, results in the cavity being strongly under-coupled ($\kappa_e \ll \kappa_i + \kappa_{MW}$), reducing the comb efficiency. This trade-off between EO comb bandwidth and efficiency limits all single-resonator EO comb sources.

A two-resonator EO comb generator can overcome this trade-off and ensures efficient energy flow into the comb cavity (Fig. 1d). In this case, a critical coupling between the bus waveguide and the device can be achieved under the existence of the strong $\kappa_{MW}$, if the following condition, referred to as the generalized critical coupling (GCC) condition, is met (Figs. 1e and 1f):

$$\kappa_{e1} = \kappa_{i1} + \frac{4\mu^2}{\kappa_{e2} + \kappa_{i2} + \kappa_{MW}}$$

where $\kappa_{e1}$ and $\kappa_{e2}$ are the waveguide coupling rates between the bus waveguide and cavity 1 and between the output waveguide and cavity 2, respectively, $\kappa_{i1}$ and $\kappa_{i2}$ are the intrinsic loss rates for cavities 1 and 2, respectively, and $\mu$ is the coupling rate between cavities 1 and 2. The term $\frac{4\mu^2}{\kappa_{e2}+\kappa_{i2}+\kappa_{MW}} \equiv \kappa_{1eff}$ can be interpreted as an effective loss rate of cavity 1 that is induced by cavity 2, the microwave modulation, and the output waveguide (drop port). With the expression for $\kappa_{MW}$ (Eq. (1)), the two-resonator EO comb generator, which involves hundreds of frequency modes, can be simplified to a two-level system that can be solved for analytically. See detailed discussion on the expression and maximum theoretical limitation of efficiency in the Supplementary Materials.

To experimentally demonstrate the two-resonator comb generator, we fabricated devices on TFLN-on-insulator (Fig. 2a). The small ring resonator and the long racetrack resonator are used as cavities 1 and 2, respectively. A microwave signal is sent to the electrode of cavity 2 to provide phase modulation. A thermal heater is implemented in cavity 1 for efficient resonance tuning.

We first demonstrate high-efficiency and broadband EO frequency comb generation. Continuous-wave light at 1605 nm is fed into the device through the bus waveguide. A 30.925 GHz microwave signal, which matches the FSR of cavity 2, is used to drive the electrode. By applying microwave modulation, the transmission of the through-port changes from over-coupled to nearly critically coupled (inset of Fig. 2c), indicating efficient flow of pump power into the device. The frequency comb is collected at the drop port of the device (Fig. 1d) and measured by an optical spectrum analyzer and photodetectors. The output comb power increases linearly with the pump power (Fig. 2b), indicating a pump-to-comb conversion efficiency of $\eta \equiv \frac{P_{comb}}{P_{pump}} = 30\%$, where $P_{comb}$ and $P_{pump}$ are the powers in the output and input waveguides, respectively. Using 2 mW of pump power (same pump power used in Ref. [21]), the comb spans 132 nm at a -70 dBm power level and has efficiency $\eta = 30\%$ (Fig. 2c). Compared to state-of-the-art integrated EO combs [21], our method yields two orders of magnitude improvement in conversion efficiency and 2.2 times broader bandwidth measured at the -70 dBm power level with the same pump power (Fig. 2c). It should be noted that in the single-resonator TFLN EO comb generator [21], the pump intensity is 40 dB higher than the first comb line due to the low efficiency from the strong under-coupling of the resonator (purple trace in Fig. 2c). For our two-resonator device, the output spectrum in the through port (residual pump, Fig. 3a) is discussed below.

Cavity 1 has an infinite number of frequency modes that can lead to mode-crossing with the modes of cavity 2, potentially causing significant comb power loss. To overcome this, we use an optimization algorithm to design the FSRs of cavity 1 and 2 so that they have overlapping resonances at the pump frequency only, and mis-aligned resonances for most other frequencies (Vernier effect, Fig. 3a). To verify this, we collect the generated combs from both the through port and drop port. The comb profile measured at the through port clearly shows the optimized Vernier effect (top panel of Fig. 3b) and the spectrum at the drop port preserves the linear slope (Fig. 3b, bottom panel) without significant cut-off [24], indicating that the mode-crossing loss is minimized. The profile of the comb collected at the through port is the result of the comb generation inside cavity 2 followed by "cavity filtering" in cavity 1. This signal can also be useful for, e.g., heterodyne measurements, as local oscillators, or clock references.

Next, enabled by the high efficiency and broad bandwidth, we demonstrate that the device can serve as an on-chip ultra-fast pulse source, which is important for nonlinear applications such as broadband parametric frequency conversion and optical atomic clocks. To characterize the time-domain signal of the output frequency comb, the optical output is sent to an erbium-doped fiber amplifier (EDFA) and dispersion-compensating fiber followed by a second-harmonic generation-based intensity autocorrelator (See Fig. 4a for the setup). The full-width-at-half-maximum (FWHM) of the autocorrelator trace is 1.072 ps, corresponding to a pulse FWHM of 536 fs (Fig. 4b). We infer that the pulse FWHM in the output facet of the chip is around 336 fs after extracting the total dispersion of the fiber output path of 36 fs/nm. The two sidelobes near the Lorentzian-shaped pulse are caused by the non-uniform gain coefficient of our EDFA (see Supplementary Materials).

Finally, we demonstrate the observation of EO and $\chi^{(3)}$ combined high-power frequency combs in a single device, enabled by the formation of ultra-fast pulses with high circulating peak power inside the resonator. To demonstrate this effect, a 63 mW pump power (in the bus waveguide) is used to feed the device (see the setup in Fig. 4a) and the output frequency comb features a 32% conversion efficiency, 20 mW on-chip comb power, and broadened bandwidth of 161 nm (Fig. 4d). As a result, we infer that an estimated ~85 W peak pulse power (~1 W average power) is circulating inside the comb resonator (cavity 2) (see Supplementary Materials), large enough to stimulate the additional Raman [28] and four-wave mixing effects [9] (Fig. 4e). Optimizing the dispersion of the current device from normal dispersion to anomalous dispersion could further enhance the four-wave mixing effect, which could be useful for ultrabroad comb generation. Combining $\chi^{(2)}$ and $\chi^{(3)}$ nonlinearities also provides an intriguing opportunity for investigating new regimes of nonlinear optical dynamics such as the proposed "band soliton" [29], and might enable superior OFCs, owing to the ability to reach both the ultra-fast and high-power regimes in the same resonator under EO modulation.

In summary, we demonstrate high-efficiency and broadband electro-optic frequency combs using a coupled-resonator structure. We show that it can be used as an integrated femtosecond pulse source and can stimulate combined second- and third- order nonlinear process in the ultra-fast high-power regime. In addition, we provide a theoretical model that simplifies the coupled-resonator system supporting hundreds of energy levels to a two-level system. We use this formalism to show that the comb efficiency can be improved and ultimately reach the fundamental limit $\xi = \frac{\kappa_{e2}}{\kappa_{e2}+\kappa_{i2}}$ by further optimizing the GCC condition, intrinsic $Q$, and the mode-crossing effect (see Supplementary Materials). Simultaneously achieving high efficiency and broad bandwidth can enable previous state-of-the-art integrated EO combs [21] to address a broad range of applications. For example, an already demonstrated 100-fold improvement in comb efficiency can lead to a 20 dB increase in the signal-to-noise ratio of frequency-multiplexed applications such as optical communications [1]. These advances also allow combs to be generated with much lower pump powers for energy-efficient optical neural networks [5,6]. Furthermore, the ability to generate femtosecond pulses on chip is important for nonlinear photonics, optical atomic clocks, optical sensing, and time-bin encoded optical computing. Finally, the high conversion efficiency of our device opens the door for a generation of broad EO combs for entangled photons, broadly enabling frequency-domain quantum information processing [26,30].

**References:**

[1] P. Marin-Palomo, J. N. Kemal, M. Karpov, A. Kordts, J. Pfeifle, M. H. P. Pfeiffer, P. Trocha, S. Wolf, V. Brasch, M. H. Anderson, R. Rosenberger, K. Vijayan, W. Freude, T. J. Kippenberg, and C. Koos, *Microresonator-Based Solitons for Massively Parallel Coherent Optical Communications*, Nature **546**, 274 (2017).

[2] M. G. Suh and K. J. Vahala, *Soliton Microcomb Range Measurement*, Science (80-. ). **359**, 884 (2018).

[3] N. Picqué and T. W. Hänsch, *Frequency Comb Spectroscopy*, Nat. Photonics **13**, 146 (2019).

[4] S. B. Papp, K. Beha, P. Del'Haye, F. Quinlan, H. Lee, K. J. Vahala, and S. A. Diddams, *Microresonator Frequency Comb Optical Clock*, Optica **1**, 10 (2014).

[5] X. Xu, M. Tan, B. Corcoran, J. Wu, A. Boes, T. G. Nguyen, S. T. Chu, B. E. Little, D. G. Hicks, R. Morandotti, A. Mitchell, and D. J. Moss, *11 TeraFLOPs per Second Photonic Convolutional Accelerator for Deep Learning Optical Neural Networks*, Nature **589**, 21 (2020).

[6] J. Feldmann, N. Youngblood, M. Karpov, H. Gehring, X. Li, M. Stappers, M. Le Gallo, X. Fu, A. Lukashchuk, A. S. Raja, J. Liu, C. D. Wright, A. Sebastian, T. J. Kippenberg, W. H. P. Pernice, and H. Bhaskaran, *Parallel Convolutional Processing Using an Integrated Photonic Tensor Core*, Nature **589**, 52 (2021).

[7] M. Kues, C. Reimer, J. M. Lukens, W. J. Munro, A. M. Weiner, D. J. Moss, and R. Morandotti, *Quantum Optical Microcombs*, Nat. Photonics **13**, 170 (2019).

[8] J. M. Lukens and P. Lougovski, *Frequency-Encoded Photonic Qubits for Scalable Quantum Information Processing*, Optica **4**, 8 (2016).

[9] T. J. Kippenberg, A. L. Gaeta, M. Lipson, and M. L. Gorodetsky, *Dissipative Kerr Solitons in Optical Microresonators*, Science (80-. ). **361**, (2018).

[10] C. Bao, L. Zhang, A. Matsko, Y. Yan, Z. Zhao, G. Xie, A. M. Agarwal, L. C. Kimerling, J. Michel, L. Maleki, and A. E. Willner, *Nonlinear Conversion Efficiency in Kerr Frequency Comb Generation*, Opt. Lett. **39**, 6126 (2014).

[11] X. Xue, X. Zheng, and B. Zhou, *Super-Efficient Temporal Solitons in Mutually Coupled Optical Cavities*, Nat. Photonics **13**, 616 (2019).

[12] X. Xue, Y. Xuan, Y. Liu, P. H. Wang, S. Chen, J. Wang, D. E. Leaird, M. Qi, and A. M. Weiner, *Mode-Locked Dark Pulse Kerr Combs in Normal-Dispersion Microresonators*, Nat. Photonics **9**, 594 (2015).

[13] B. Y. Kim, Y. Okawachi, J. K. Jang, M. Yu, X. Ji, Y. Zhao, C. Joshi, M. Lipson, and A. L. Gaeta, *Turn-Key, High-Efficiency Kerr Comb Source*, Opt. Lett. **44**, 4475 (2019).

[14] Ó. B. Helgason, F. R. Arteaga-Sierra, Z. Ye, K. Twayana, P. A. Andrekson, M. Karlsson, J. Schröder, and Victor Torres-Company, *Dissipative Solitons in Photonic Molecules*, Nat. Photonics **15**, 305 (2021).

[15] T. Sakamoto, T. Kawanishi, and M. Izutsu, *Asymptotic Formalism for Ultraflat Optical Frequency Comb Generation Using a Mach-Zehnder Modulator*, Opt. Lett. **32**, 1515

(2007).

[16] S. Ozharar, F. Quinlan, I. Ozdur, S. Gee, and P. J. Delfyett, *Ultraflat Optical Comb Generation by Phase-Only Modulation of Continuous-Wave Light*, IEEE Photonics Technol. Lett. **20**, 36 (2008).

[17] M. Kourogi, N. Ken'ichi, and M. Ohtsu, *Wide-Span Optical Frequency Comb Generator for Accurate Optical Frequency Difference Measurement*, IEEE J. Quantum Electron. **29**, 2693 (1993).

[18] K.-P. Ho and J. M. Kahn, *Optical Frequency Comb Generator Using Phase Modulation in Amplified Circulating Loop*, IEEE Photonics Technol. Lett. **5**, 721 (1993).

[19] S. Xiao, L. Hollberg, N. R. Newbury, and S. A. Diddams, *Toward a Low-Jitter 10 GHz Pulsed Source with an Optical Frequency Comb Generator*, Opt. Express **16**, 8498 (2008).

[20] A. Rueda, F. Sedlmeir, M. Kumari, G. Leuchs, and H. G. L. Schwefel, *Resonant Electro-Optic Frequency Comb*, Nature **568**, 378 (2019).

[21] M. Zhang, B. Buscaino, C. Wang, A. Shams-Ansari, C. Reimer, R. Zhu, J. M. Kahn, and M. Lončar, *Broadband Electro-Optic Frequency Comb Generation in a Lithium Niobate Microring Resonator*, Nature **568**, 373 (2019).

[22] A. Shams-Ansari, M. Yu, Z. Chen, C. Reimer, M. Zhang, N. Picqué, and M. Lončar, *An Integrated Lithium-Niobate Electro-Optic Platform for Spectrally Tailored Dual-Comb Spectroscopy*, ArXiv **2003.04533**, (2020).

[23] D. Zhu, L. Shao, M. Yu, R. Cheng, B. Desiatov, C. Xin, Y. Hu, J. Holzgrafe, S. Ghosh, A. Shams-Ansari, E. Puma, N. Sinclair, C. Reimer, M. Zhang, and M. Loncar, *Integrated Photonics on Thin-Film Lithium Niobate*, Adv. Opt. Photonics **13**, (2021).

[24] B. Buscaino, M. Zhang, M. Lončar, and J. M. Kahn, *Design of Efficient Resonator-Enhanced Electro-Optic Frequency Comb Generators*, J. Light. Technol. **38**, 1400 (2020).

[25] M. Kourogi, T. Enami, and M. Ohtsu, *A Coupled-Cavity Monolithic Optical Frequency Comb Generator*, IEEE Photonics Technol. Lett. **8**, 1698 (1996).

[26] Y. Hu, C. Reimer, A. Shams-Ansari, M. Zhang, and M. Loncar, *Realization of High-Dimensional Frequency Crystals in Electro-Optic Microcombs*, Optica **7**, 1189 (2020).

[27] Y. Hu, M. Yu, D. Zhu, N. Sinclair, A. Shams-Ansari, L. Shao, J. Holzgrafe, E. Puma, M. Zhang, and M. Loncar, *Reconfigurable Electro-Optic Frequency Shifter*, ArXiv **2005.09621**, (2020).

[28] M. Yu, Y. Okawachi, R. Cheng, C. Wang, M. Zhang, A. L. Gaeta, and M. Lončar, *Raman Lasing and Soliton Mode-Locking in Lithium Niobate Microresonators*, Light Sci. Appl. **9**, (2020).

[29] A. K. Tusnin, A. M. Tikan, and T. J. Kippenberg, *Nonlinear States and Dynamics in a Synthetic Frequency Dimension*, Phys. Rev. A **102**, 1 (2020).

[30] P. Imany, N. B. Lingaraju, M. S. Alshaykh, D. E. Leaird, and A. M. Weiner, *Probing Quantum Walks through Coherent Control of High-Dimensionally Entangled Photons*, Sci. Adv. **6**, eaba8066 (2020).

[31] S. Ramelow, A. Farsi, S. Clemmen, J. S. Levy, A. R. Johnson, Y. Okawachi, M. R. E. Lamont, M. Lipson, and A. L. Gaeta, *Strong Polarization Mode Coupling in Microresonators*, Opt. Lett. **39**, 5134 (2014).

**Acknowledgments:** We thank Cheng Wang for helpful discussion. This work is supported by AFOSR grants FA9550-19-1-0376 and FA9550-19-1-0310, DARPA LUMOS HR0011-20-C-137, NASA 80NSSC21C0583, AFRL FA9550-21-1-0056, NSF ECCS-1839197, ARO W911NF2010248, DOE DE-SC0020376, Harvard Quantum Initiative, Facebook, Maxim Integrated (now Analog Devices), Inphi (now Marvell) and National Science Foundation under Grant ECCS-1740291 E2CDA. Device fabrication was performed at the Harvard University Center for Nanoscale Systems.

**Competing interests:** M.Z. and M.L. are involved in developing lithium niobate technologies at HyperLight Corporation.

**Disclaimer:** The views, opinions and/or findings expressed are those of the author and should not be interpreted as representing the official views or policies of the Department of Defense or the U.S. Government.

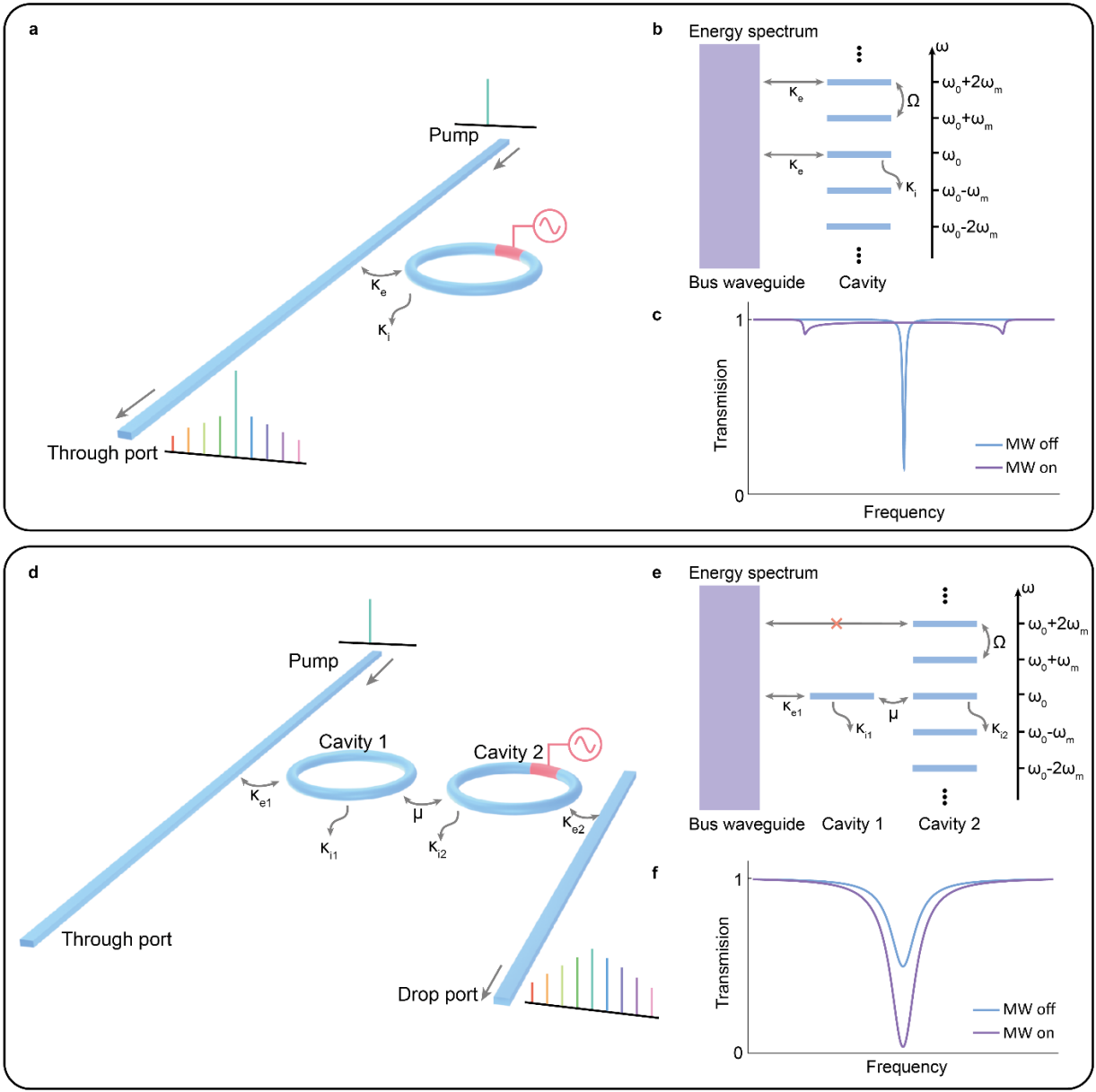

**Fig. 1. Concept of the two-resonator EO comb and generalized critical coupling condition. a, b, c,** Schematics of the device structure (a), energy-level description (b), and transmission spectrum (c) of the single-resonator EO frequency comb generator. EO modulation creates efficient coupling $\Omega$ between adjacent frequency modes of the cavity (b). All the frequency modes are coupled to the input bus waveguide. The rate $\kappa_e$ is the waveguide-cavity coupling rate. When the microwave signal is off, the pump resonance is nearly critically coupled but becomes strongly under-coupled when the microwave signal is turned on. **d, e, f,** Schematics of the device structure (d), energy-level description (e), and through-port transmission spectrum (f) of the two-resonator EO frequency comb generator. The inclusion of cavity 1 effectively causes an over-coupling between the pump mode of cavity 2 and the bus waveguide, while rejecting other frequency modes of cavity 2 (e). Such coupled-ring structures can achieve a general critical coupling condition, resulting in a high extinction ratio of the pump resonance when the microwave signal is turned on

(f). The rate $\kappa_{e1}$ ($\kappa_{e2}$) is the coupling rate between waveguide and cavity 1 (2). $\mu$ is the coupling rate between cavity 1 and 2. $\kappa_{i1}$ and $\kappa_{i2}$ are the intrinsic loss rates of cavity 1 and cavity 2, respectively. MW, microwave.

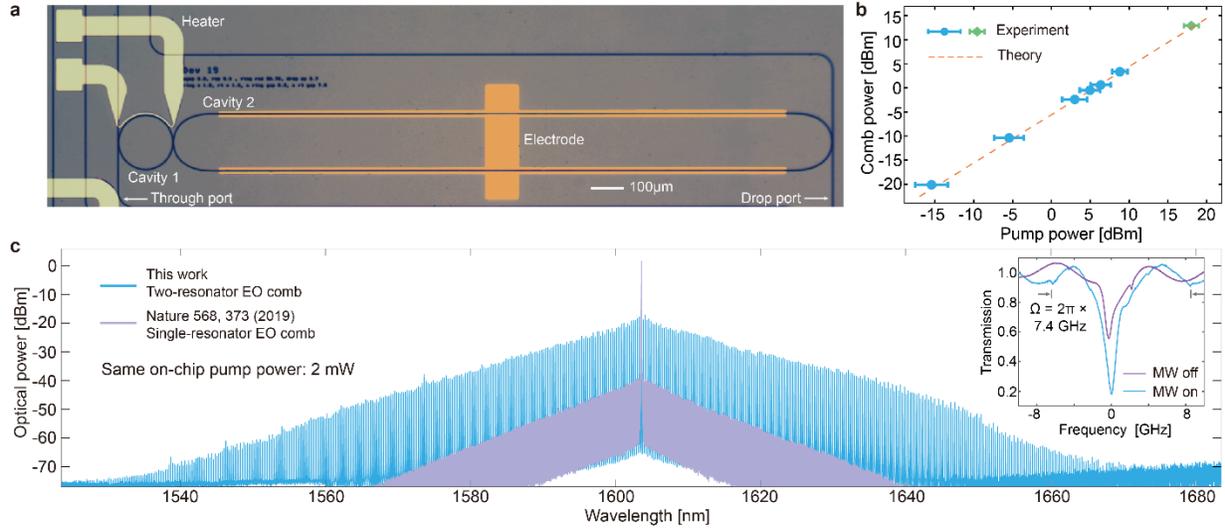

**Fig. 2. High-efficiency and broadband EO comb generator. a**, Device optical image (false color). Cavity 1 is controlled using a thermal heater (light yellow) and cavity 2 is modulated by a microwave signal applied to gold electrodes (orange). **b,** Measured comb power vs. pump power, indicating a pump-to-comb conversion efficiency of 30%. The data points labelled by blue circle and green diamond represent data from two different devices with the same parameters. Data points with blue circles represent similar comb shapes with the comb in (c), while green diamonds represent the nonlinear comb in the ultra-fast high-power regime (see Fig. 4d). **c,** Optical spectra of the two-resonator EO comb generator (this work) and single-resonator structure [21]. The measured spectrum is offset by the output facet loss in order to compare the two on-chip comb spectra at the same on-chip pump power (2 mW). At a -70 dBm power level, the bandwidth of the two-resonator comb is 132 nm, while the bandwidth of the single-resonator EO comb [21] is 60 nm. Insets shows the transmission spectrum of the through port (see Fig. 1d) when the microwave signal is off (purple) and on (blue). The span between two shallow transmission dips (vertical lines with arrows) is the conventional resonance broadening of the EO comb generator, which gives the coupling rate $\Omega = 2\pi \times 7.4$ GHz as well as the modulation index $\beta = 2\pi \frac{\Omega}{FSR} = 2\pi \frac{2\pi \times 7.4\ GHz}{2\pi \times 30.925\ GHz} = 0.48\pi$.

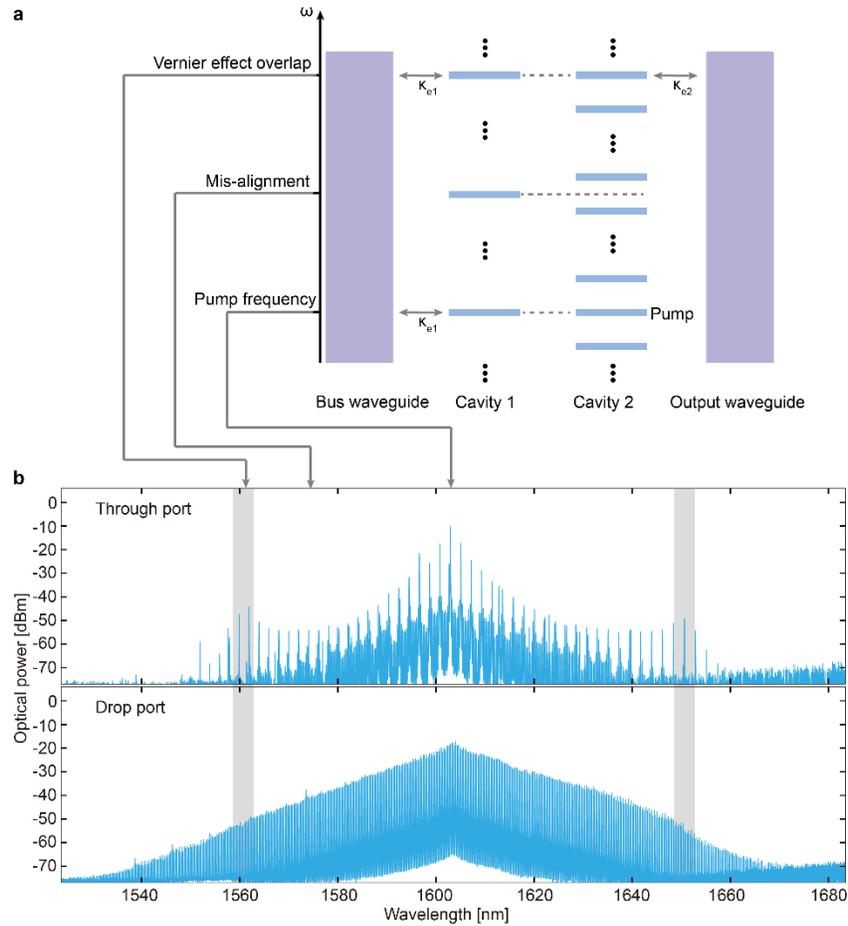

**Fig. 3. Limitation and optimization of the mode-crossing effect. a,** Illustration of the Vernier effect for optimizing mode-crossing-induced loss using an energy level diagram. Cavity 1 is designed to have its resonance aligned with the pump frequency of cavity 2 and is misaligned across a wide range of other frequencies. This overcomes the losses that would otherwise be induced by mode-crossing. The through port comb line will have higher power close to the overlap area (grey shaded region in (b)) and vice versa. **b,** Output spectrum from the through port and drop port of the device shown in Fig. 2c. The shape of the through port spectrum reflects how the Vernier effect of the cavity resonances of two rings affects the output comb in both the through port and drop port. The final optimized device FSRs are 254.54 GHz (cavity 1) and 30.925 GHz (cavity 2).

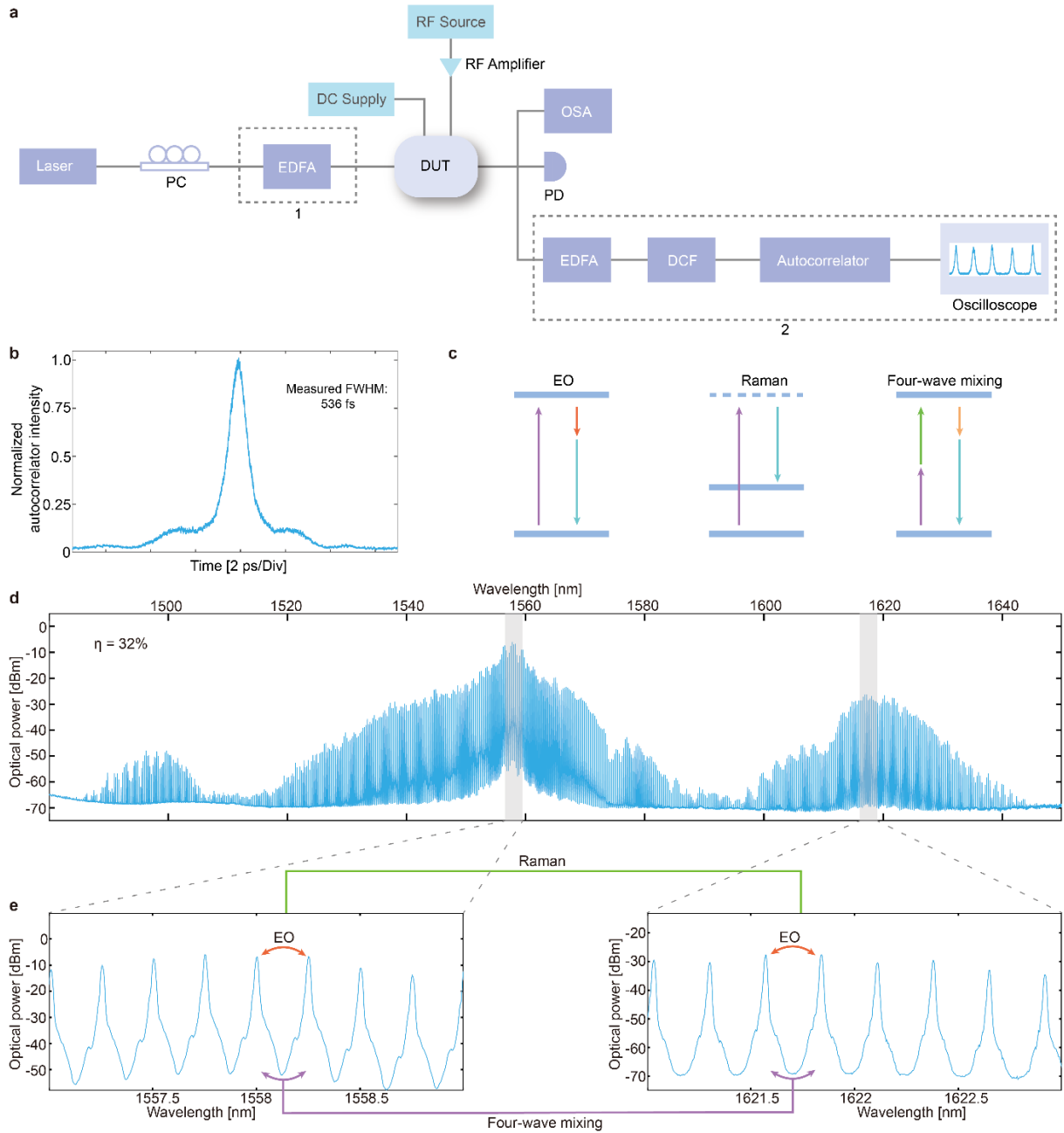

**Fig. 4. Femtosecond pulse source and multiple-combined nonlinearity in ultra-fast high-power regime. a,** Measurement setup. To characterize generated pulses in the time domain, the dashed box 2 is used. Dashed box 1 is used only when evaluating the high-power performance of the device. DUT, device under test; EDFA, Erbium-doped fiber amplifier; DCF, dispersion-compensating fiber; OSA, optical spectrum analyzer; PD, photodetector. **b,** Time-domain trace of the output comb measured by an autocorrelator. The measured pulse FWHM is 536 fs, indicating a FWHM of 336 fs in the output facet of the chip after extracting the total dispersion of the fiber output path of 36 fs/nm. **c,** Illustration of the EO, Raman, and four-wave mixing effect. **d,** Device output spectrum contains both the EO and $\chi^{(3)}$ effect when pumped with high input power. The bandwidth is broadened to 161 nm due to Raman and four-wave mixing and the conversion efficiency is 32%. **e,** Illustration of the contained EO, Raman, and four-wave mixing nonlinear processes.

# Supplementary Materials

**Device fabrication.**
Our devices are fabricated from a commercial x-cut lithium niobate (LN) on insulator wafer (NANOLN), with a 600 nm LN layer, 2 µm buried oxide (thermally grown), on a 500 µm Si handle. Electron-beam lithography with hydrogen silsesquioxane (HSQ) resist followed by Ar$^+$-based reactive ion etching (350 nm etch depth) are used to pattern the optical layer of the devices, including the rib waveguides and micro-ring resonators. The devices are cleaned, and microwave electrodes (15 nm of Ti, 300 nm of Au) are defined by photolithography followed by electron-beam evaporation and a bilayer lift-off process. One layer of SiO$_2$ (800 nm) using plasma-enhanced chemical vapor deposition (PECVD) is used to clad the devices. The heater (15 nm of Ti and 200 nm Pt) for Ring 1 is defined by photolithography followed by electron-beam evaporation and lift-off. The heater is designed as a short metal strip (5 µm width) that is placed 3 µm away from the resonator on top of the SiO$_2$ cladding. The resistance of the heater is ∼140 Ω (including parasitic resistance from routing strips). Tuning of the ring resonance by the FSR is achieved using a current of ∼50 mA.

**Frequency comb characterization.**
The measurement setup is illustrated in Fig. S1. Telecommunication-wavelength light from a fiber-coupled tunable laser (SANTEC TSL-510) passes through a polarization controller and is coupled to the LN chip using a lensed fiber. The output is collected using a lensed fiber and then sent to an optical spectrum analyzer (OSA) with a spectral resolution of 0.02 nm for characterization of the frequency comb. The microwave signal is generated from a synthesizer followed by a microwave amplifier and delivered to the electrodes on a device using an electrical probe. The microwave driving power is 2.2 W.

The circulating power in Fig. 4d is inferred from the comb power $P_{comb} = 20$ mW in the bus waveguide, which gives an intra-cavity power of cavity 2: $P_{intra} = P_{comb} \times \frac{Finesse}{\pi}$. The $Finesse$ is calculated by $Finesse = \frac{\kappa_2}{FSR_2} = \frac{222 \text{ MHz}}{30.925 \text{ GHz}} = 139.3$. Therefore $P_{intra} = 0.9$ W and the peak power of the pulse can be inferred by $P_{peak} = \frac{t_{RT}}{\tau} P_{intra} = 85$ W in which $t_{RT} = 32$ ps and $\tau = 336$ fs are the roundtrip time of cavity 2 and the pulse duration, respectively.

The device parameters are extracted based on the measurement on the transmission spectrums (Fig. S4 and Table S1). The parameters of cavity 1 ($\kappa_{e1}$ and $\kappa_{i1}$) and cavity 2 ($\kappa_{e2}$ and $\kappa_{i2}$) are obtained from the linewidth and extinction ratio of the resonances. The coupling $\mu$ between the two cavities is extracted by tuning the two resonances of the cavities to a degenerate point and measuring the mode-splitting, which gives a $\mu = \sqrt{\left(\frac{\text{Splitting}}{2}\right)^2 + \left(\frac{\mu_{EP}}{2}\right)^2}$ where $\mu_{EP} = \frac{\kappa_1 - \kappa_2}{2}$ is the coupling strength to reach the exceptional point of the system.

## Theoretical analysis of the two-resonator EO comb generator.
### Effective loss rate induced by microwave for EO comb.

To obtain the effective loss rate that is induced by microwave modulation, we consider the case that our cavity 2 (comb cavity) is driven by a continuous microwave signal without the existence of the cavity 1. Then Hamiltonian of the system follows a single-resonator EO comb model (we set $\hbar = 1$):

$$H = \sum_{j=-N}^{N} \omega_j a_j^\dagger a_j + \Omega \cos \omega_m t \, (a_j^\dagger a_{j+1} + h.c.)$$

in which $\omega_j$ represents the frequency of each frequency mode, $\Omega$ is the coupling rate due to microwave modulation, and $\omega_m$ is the frequency of microwave signal. We also assume the frequency modes range from $j = -N$ to $j = N$. Implementing the Heisenberg-Langevin equation gives a set of equations of motion for each mode $a_j$:

$$\dot{a}_j = \left(-i\omega_j - \frac{\kappa_2}{2}\right) a_j - i\Omega \cos \omega_m t \, (a_{j+1} + a_{j-1}) - \sqrt{\kappa_{e2}} \alpha_{in} e^{-i\omega_L t} \delta_{j,0}$$

in which $\kappa_{e2}, \kappa_{i2}$, and $\kappa_2 = \kappa_{e2} + \kappa_{i2}$ are the coupling rate between cavity 2 and output waveguide, intrinsic loss rate of the cavity 2, and total loss rate of $a_j$, respectively. The pump power and frequency are denoted by $\alpha_{in}$ and $\omega_L$, respectively. We use the Kronecker delta function $\delta_{j,0}$ to indicate that only the 0$^{th}$ mode is pumped. Therefore, we write $\omega_j$ as $\omega_j = \omega_0 + j \times FSR$ in which $\omega_0$ is the resonance frequency of the 0$^{th}$ mode. By changing rotating frames for each mode $a_j \to a_j e^{-i\omega_L t} e^{-ij\omega_m t}$, we obtain the simplified equations of motion:

$$\dot{a}_j = \left(i\Delta + i\delta - \frac{\kappa_2}{2}\right) a_j - i\frac{\Omega}{2}(a_{j+1} + a_{j-1}) - \sqrt{\kappa_{e2}} \alpha_{in} \delta_{j,0}$$

in which $\Delta = \omega_L - \omega_0$ and $\delta = \omega_m - FSR$ are the laser detuning and microwave detuning, respectively.

The steady state of such system can be analytically solved. Considering the case that $\Delta = \delta = 0$, the equations of motion become

$$0 = -\frac{\kappa_2}{2} a_j - i\frac{\Omega}{2} a_{j+1} - i\frac{\Omega}{2} a_{j-1} - \sqrt{\kappa_{e2}} \alpha_{in} \delta_{j,0}$$

Note that the equation of motion for the mode with largest mode number is

$$0 = \left(-\frac{\kappa_2}{2}\right) a_N - i\frac{\Omega}{2} a_{N-1}$$

which gives a relation $a_N = -i\frac{\Omega}{\kappa_2} a_{N-1}$. As a result, the equation for $a_{N-1}$ becomes

$$0 = \left(-\frac{\kappa_2}{2}\right) a_{N-1} - i\frac{\Omega}{2} a_{N-2} - \frac{\Omega}{2} \frac{\Omega}{\kappa_2} a_{N-1}$$

which leads to another relation $a_{N-1} = -i\frac{\Omega}{\kappa_2 \left(1+\frac{\Omega^2}{\kappa_2^2}\right)} a_{N-2}$. By iterating the above steps, we obtain the relation for an arbitrary mode

$$a_l = -i\frac{\Omega}{\kappa_2}\frac{1}{1+\cfrac{\frac{\Omega^2}{\kappa_2^2}}{1+\cfrac{\frac{\Omega^2}{\kappa_2^2}}{1+\cfrac{\kappa_2^2}{\cdots}}}}a_{l-1}$$

with the total number of the factor $\frac{\Omega^2}{\kappa_2^2}$ is $N-l$. As a result, the equation of motion for the $0^{\text{th}}$ mode is

$$0 = -\frac{\kappa_2}{2}a_0 - i\frac{\Omega}{2}\left(-i\frac{\Omega}{\kappa_2}\frac{1}{1+\cfrac{\frac{\Omega^2}{\kappa_2^2}}{1+\cfrac{\frac{\Omega^2}{\kappa_2^2}}{1+\cfrac{\kappa_2^2}{\cdots}}}}\right)a_0 - i\frac{\Omega}{2}\left(-i\frac{\Omega}{\kappa_2}\frac{1}{1+\cfrac{\frac{\Omega^2}{\kappa_2^2}}{1+\cfrac{\frac{\Omega^2}{\kappa_2^2}}{1+\cfrac{\kappa_2^2}{\cdots}}}}\right)a_0 - \sqrt{\kappa_{e2}}\alpha_{in}$$

Simplifying this equation gives

$$0 = \left(-\frac{\kappa_2}{2} - \frac{\kappa_{MW}}{2}\right)a_0 - \sqrt{\kappa_{e2}}\alpha_{in}$$

in which $\kappa_{MW} = \kappa_2 \times 2(f_n - 1)$ with

$$f_n = 1 + \cfrac{\frac{\Omega^2}{\kappa_2^2}}{1+\cfrac{\frac{\Omega^2}{\kappa_2^2}}{1+\cfrac{\frac{\Omega^2}{\kappa_2^2}}{1+\cfrac{\kappa_2^2}{\cdots}}}}$$

In the limit of N is a very large number, we can use the limit of $\lim\limits_{n\to\infty} f_n = f$ and $f = 1 + \frac{\frac{\Omega^2}{\kappa_2^2}}{f}$ to solve the final expression for $\kappa_{MW}$

$$\kappa_{MW} = \kappa_2\left(\sqrt{1+\frac{4\Omega^2}{\kappa_2^2}} - 1\right)$$

**Conversion efficiency analysis using the generalized critical coupling condition**

With the expression of $\kappa_{MW}$, the single-resonator EO comb system can be simplified as a single mode which has three loss rates $\kappa_{e2}, \kappa_{i2}$ and $\kappa_{MW}$. Therefore, the steady-state equations for a two-resonator EO comb are

$$0 = -\frac{\kappa_2 + \kappa_{MW}}{2} a_0 - i\mu d$$

$$0 = -\frac{\kappa_1}{2} d - i\mu a_0 - \sqrt{\kappa_{e1}} \alpha_{in}$$

in which $d$ is the pump mode of cavity 1, $\mu$ is the evanescent coupling rate between cavity 1 and 2, and $\kappa_{e1}, \kappa_{i1}$, and $\kappa_1 = \kappa_{e1} + \kappa_{i1}$ are the coupling rate between cavity 1 and input waveguide, intrinsic loss rate of the cavity 1, and total loss rate of $d$, respectively. Note that in the above equation, cavity 1 is pumped instead of cavity 2. Hence, the effective loss rate $\kappa_{1eff}$ for cavity 1 that is induced by the cavity 2 and the output waveguide of cavity 2 is

$$\kappa_{1eff} = \frac{4\mu^2}{\kappa_2 + \kappa_{MW}}$$

and a critical coupling condition occurs when $\kappa_{e1} = \kappa_{i1} + \frac{4\mu^2}{\kappa_2 + \kappa_{MW}}$. The output in the through port waveguide is $d_{out} = \alpha_{in} + \sqrt{\kappa_{e1}} d = \alpha_{in} \left( \frac{\kappa_{1eff} + \kappa_{i1} - \kappa_{e1}}{\kappa_{1eff} + \kappa_{i1} + \kappa_{e1}} \right)$. The amplitude that is lost in the intrinsic loss of cavity 1 is $d_{int} = \sqrt{\kappa_{i1}} d = \alpha_{in} \frac{2\sqrt{\kappa_{e1}\kappa_{i1}}}{\kappa_{1eff} + \kappa_{i1} + \kappa_{e1}}$. Therefore, the power that flows to cavity 2 can be obtained as

$$P_2 = (|\alpha_{in}|^2 - |d_{out}|^2 - |d_{int}|^2)$$

Finally, the output comb power that can be collected is dominated by the factor $\xi = \frac{\kappa_{e2}}{\kappa_{e2} + \kappa_{i2}}$ which quantifies the energy going to the output comb channel when light is circulating inside cavity 2. Note that the factor $\xi$ does not include the losses of other frequency modes except the pump mode back to cavity 1 and input waveguide, which is negligible due to the off-resonant condition between cavity 1 and cavity 2. For some frequency modes that at which cavity 1 and cavity 2 are resonant due to the Vernier effect, the loss is minor since the power of those frequency modes are much lower than the modes close to pump mode. Thus, the conversion efficiency is

$$\eta = \frac{P_{comb}}{P_{in}} = \frac{P_2}{P_{in}} \times \xi$$

Therefore the parameter $\xi$ sets the fundamental limit of the conversion efficiency. For example, in this work, cavity 2 is nearly critically coupled to the output comb waveguide, leading to a ~50% theoretical limit of the conversion efficiency. Putting the expression of $P_2$, we obtain the final efficiency as

$$\eta = \left( 1 - \left( \frac{\kappa_{1eff} + \kappa_{i1} - \kappa_{e1}}{\kappa_{1eff} + \kappa_{i1} + \kappa_{e1}} \right)^2 - \left( \frac{2\sqrt{\kappa_{e1}\kappa_{i1}}}{\kappa_{1eff} + \kappa_{i1} + \kappa_{e1}} \right)^2 \right) \times \frac{\kappa_{e2}}{\kappa_{e2} + \kappa_{i2}}$$

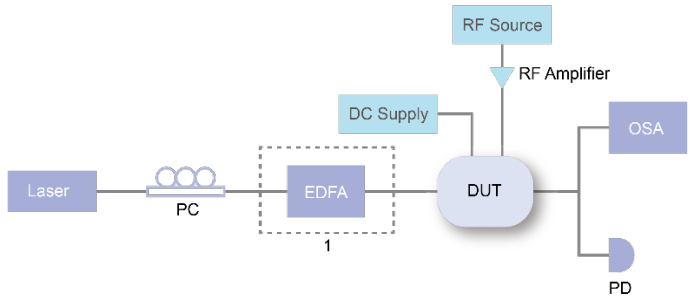

**Fig. S1. Measurement setup for Fig. 2 and 3.** The two-resonator device is characterized using the above setup. In the experiment of Fig. 2b, an EDFA is used to obtain higher pump power. In the experiment of Fig. 2c, 3b, the EDFA is not used. DUT, device under test; EDFA, Erbium-doped fiber amplifier; OSA, optical spectrum analyzer; PD, photodetector.

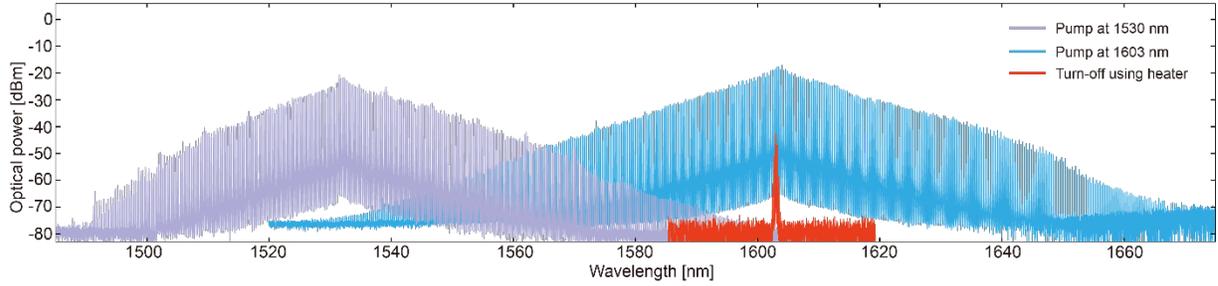

**Fig. S2. Tunability of pump frequency and turning on-off using the heater.** Spectra of the device pumped at 1530 nm (purple trace) and 1603 nm (blue trace and red trace). By tuning the resonances of cavity 1 to match the resonances of cavity 2, the device can be pumped at different wavelengths. The cut-off at wavelengths that are far blue-detuned (purple trace) is due to TE-TM (transverse electric-transverse magnetic) polarization crossing, which affects the TE-designed cavity 2, and can be minimized by additional dispersion engineering [31]. The comb can also be turned off simply by tuning the resonance of cavity 1 to be mis-aligned from cavity 2. The blue and red traces show the on and off comb states by changing the heater voltages without changing the laser or microwave drive, showing an excellent extinction ratio.

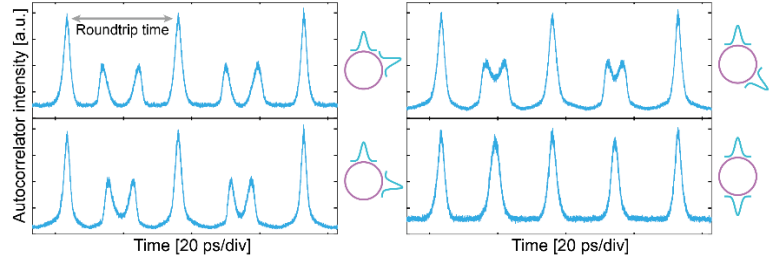

**Fig. S3. Tuning the dual-pulse in one roundtrip time.** The high conversion-efficiency allows us to measure the signal in the time-domain under varied optical detuning. Unlike dark-pulse Kerr combs or platicons in a two-resonator Kerr frequency comb generator [14], which exhibits a completely different mechanism in both spectral and temporal domains compared to their single-resonator counterparts, our two-resonator structure preserves the time-domain features of the single-resonator EO combs. The output signal shows that there are two pulses in one round-trip time and the delay between the two pulses can be tuned by changing the optical detuning, which is identical to the conventional single-resonator EO comb.

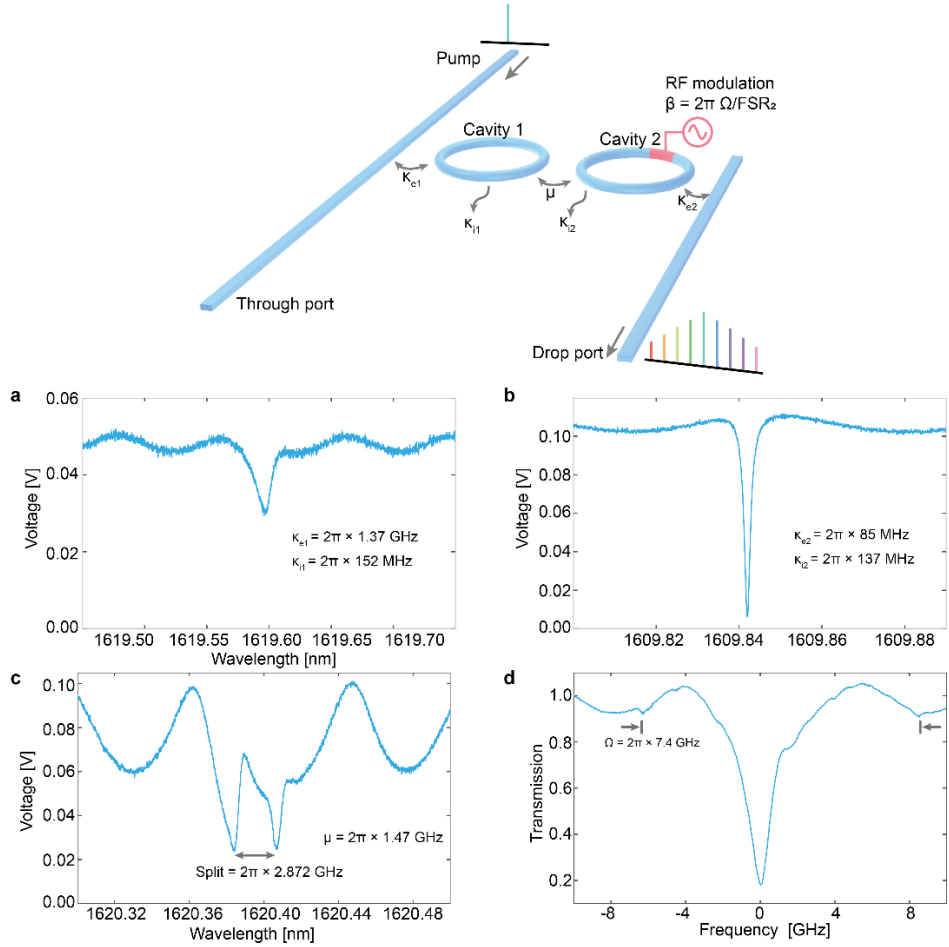

**Fig. S4. Device parameter analysis. a, b,** Transmission spectrum of a single cavity 1 (a) and 2 (b) with the same fabrication parameters as the two-resonator device. **c,** Transmission spectrum of a two-resonator device on the through port. **d,** Transmission spectrum of a two-resonator device when microwave is on. (c) and (d) are measured on two different two-resonator devices with the same fabrication parameters. The extracted parameters give a theoretical conversion efficiency of 28%.

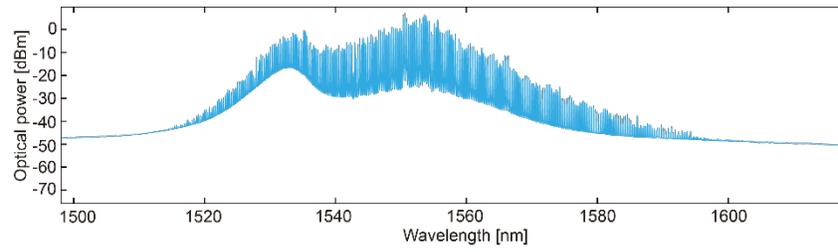

**Fig. S5. Frequency spectrum after passing the EDFA.** The spectrum shows the frequency comb after amplification by the EDFA in the time-domain pulse measurement of Fig. 4b.

Table S1. Parameters of the devices

| Parameters | Values |
| --- | --- |
| Waveguide-cavity 1 coupling $\kappa_{e1}$ | $2\pi \times 1.37$ GHz |
| Intrinsic loss rate $\kappa_{i1}$ | $2\pi \times 152$ MHz |
| Waveguide-cavity 2 coupling $\kappa_{e2}$ | $2\pi \times 85$ MHz |
| Intrinsic loss rate $\kappa_{i2}$ | $2\pi \times 137$ MHz |
| Cavity 1-cavity 2 coupling $\mu$ | $2\pi \times 1.47$ GHz |
| Mode-coupling rate $\Omega$ | $2\pi \times 7.40$ GHz |
| Modulation index $\beta$ | $0.48\pi$ |